\documentclass[letterpaper,10pt]{article}
\usepackage{times}
\usepackage{multicol}
\usepackage{lipsum} 
\usepackage{titlesec}
\usepackage{fancyhdr}
\usepackage[margin=1in]{geometry}
\usepackage{graphicx} 
\usepackage{hyperref}
\usepackage{caption}
\usepackage{tabularx} 
\usepackage{booktabs} 
\usepackage{float} 
\usepackage{stfloats}

\titleformat{\section}{\bfseries\Large}{\thesection.}{1em}{}
\titleformat{\subsection}{\bfseries\large}{\thesubsection.}{1em}{}

\begin{document}

\title{Cultural Windows: Towards a Workflow for Immersive Journeys into Global Living Spaces}
\author{Hessam Djavaherpour \\
\texttt{djavaherpour@gmail.com} \\
Inria, Bordeaux, France
\and
Pierre Dragicevic \\
\texttt{pierre.dragicevic@inria.fr} \\
Inria, Bordeaux, France
\and
Yvonne Jansen \\
\texttt{yvonne.jansen@cnrs.fr} \\
CNRS, Inria, Bordeaux, France}

\date{}

\maketitle

\begin{figure}[h!]
    \centering
    \includegraphics[width=\textwidth]{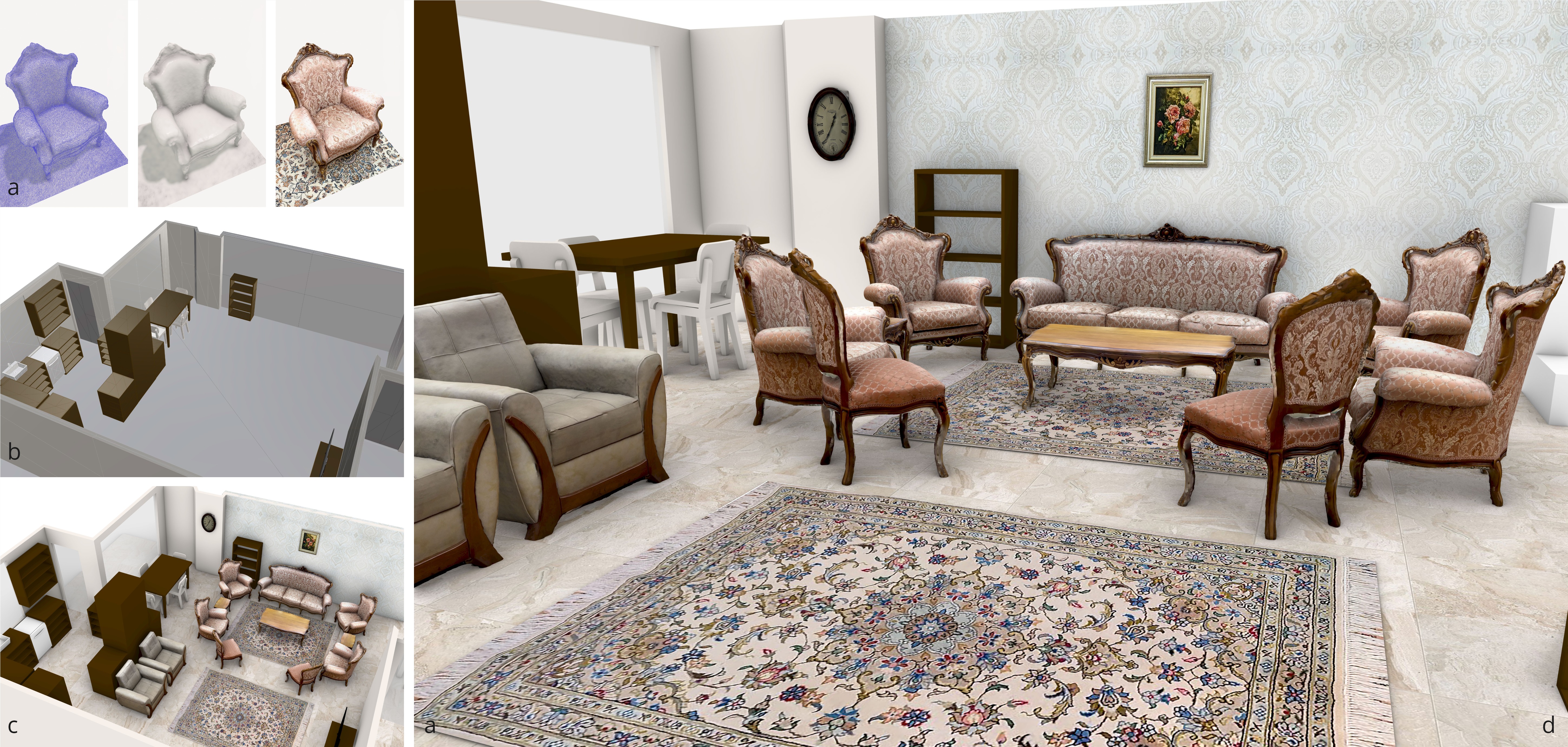}
    \caption{``Cultural Windows'' workflow for immersive visualizations of global interiors: (a) 3D scanning and processing of unique design elements, (b) Ensuring privacy with generic objects and non-photorealistic models, (c) Incorporating 3D models with original textures and materials, and (d) Preparing the final scene for VR exploration.}
    \label{fig:teaser}
\end{figure}

\begin{abstract}
    ``Cultural Windows'' is a research initiative designed to enrich cross-cultural understanding through immersive extended reality (XR) experiences. This project proposes a workflow for deploying AR and VR platforms, allowing users to explore living spaces from diverse cultures and socio-economic statuses. The process involves 3D scanning of culturally significant objects, creating accurate models of living spaces, and integrating them into immersive systems to facilitate engagement with global living designs. Targeted at individuals curious about how people live in different parts of the world, the project aims to expand cross-cultural understanding and design perspectives, providing insights into the effectiveness of immersive technologies in cultural education. By detailing its conceptual framework, ``Cultural Windows'' aims to enhance comprehension and appreciation of global domestic aesthetics by comparing participants' perceptions with immersive, realistic representations of living spaces from different cultures. This can help bridge the gap between preconceived notions and the actual appearance and feel of these spaces.
\end{abstract}

\begin{multicols}{2}

\noindent \textbf{Keywords:} Cross-Cultural Visualization, Extended Reality (XR), Cultural Awareness in Design.


\section{Introduction}
Immersive technology has been increasingly used as a tool to enable the experience of ``being there virtually,'' such as virtually traveling to unknown places \cite{tussyadiah2018virtual} or in the form of immersive news reports \cite{de2010immersive}. While virtual travel has been explored extensively, with numerous VR travel applications available on the market, these generally focus on visiting landmarks or landscapes, such as ``National Geographic Explore VR'' and ``Rebuilding Notre Dame'' \cite{meta_vacation2024}. These applications facilitate the virtual visit of known touristic destinations but rarely enable people to experience how locals live in the culture around those destinations. This aspect has been highlighted as one of the appeals of peer-to-peer accommodation platforms \cite{wildish2019living}.

Experiencing how locals live has also been the subject of a variety of non-immersive media projects. Peter Menzel, in his projects ``Material World: Global Family Portrait'' \cite{menzel2021global} and ``Hungry Planet: What the World Eats'' \cite{menzel2021daily}, visited people around the world and photographed all their belongings or the food they ate in a week. Similarly, the ``Dollar Street'' initiative by Gapminder \cite{dollarstreet2021} shared extensive sets of photographs and videos of people’s living situations and body parts. While these projects provide a rich and rare peek into other people’s everyday lives, the shared content often feels quite private and invasive, limiting the number of volunteers willing to provide such intimate views of their living situations.

In this article, we explore an alternative approach that minimizes privacy invasion while enabling the exploration and reproduction of diverse cultural living spaces through immersive experiences. 
In a similar spirit to ``Immersive Humanitarian Visualization'' \cite{dragicevic2022towards}, which proposes to use immersive technologies to make distant humanitarian issues more tangible, ``Cultural Windows'' applies these technologies to foster cross-cultural understanding. However, our project shifts the focus from human tragedies to cultural exploration. By developing a workflow using 3D scanning and extended reality (XR) for immersive journeys into living spaces from other cultures and socioeconomic statuses, we explore architectural and interior design diversity. Participants engage with virtual environments where essential design elements are detailed realistically, while ensuring that personal and sensitive aspects are omitted to protect privacy (Figure \ref{fig:teaser}).

As a preliminary evaluation of this workflow, we conducted an informal study in a focus group setting with designers and architects from diverse backgrounds, distinct from the target culture. Participants described a typical living room in a specific cultural setting, in this case, Iran, to establish a baseline for comparing their preconceptions against virtual recreations. The primary aim was to determine if such immersive experiences could promote understanding of various cultures.
Most participants observed that their understanding of the culture improved after the virtual experiences.  
Overall, participant feedback suggests that ``Cultural Windows'' is a promising approach for fostering cultural appreciation and reducing biases in both educational and professional contexts, such as interior design and set design.
``Cultural Windows'' opens up directions for enhancing global cultural understanding, and supporting a more inclusive and diverse approach across disciplines, sensitive to the cultural contexts of the communities it serves.


\section{Literature Review}

In this section, we explore the diverse scholarly landscape that informs and supports the development of ``Cultural Windows,'' defining the theoretical and practical basis of immersive technologies and their intersection with cultural understanding and design education. This review begins with understanding cultural visualization and empathy, examining cultural representation in media, empathy through photography, and empathy through immersive technology. We also discuss the specific applications of immersive technologies in educational and design contexts, illustrating their role in bridging knowledge gaps and fostering a more inclusive approach to cultural representation. Finally, we identify existing challenges and opportunities within this field, setting the stage for the contributions of the ``Cultural Windows'' project.

\subsection{Cultural Visualization and Empathy}

\textbf{Cultural Representation in Media:} Media significantly shapes public perception and understanding of ethnic and cultural groups. It influences societal attitudes by molding how individuals view their own and other ethnic groups. Jeffres et al. \cite{jeffres2011} found a strong correlation between television viewing and perceptions of cultural values, affecting societal interactions. Rigoni \& Saitta \cite{rigoni2012} discuss how media plays a crucial role in the representation and participation of identity across globalized public spaces. Additionally, media impacts cultural perceptions through its portrayal of food cultures, influencing how audiences engage with different culinary traditions \cite{tortolini2021}.\\[1em]
\textbf{Empathy Through Photography:} The use of visual media, particularly photography, enhances cross-cultural empathy and awareness. Peter Menzel's projects offer compelling visual narratives that capture diverse cultural backgrounds. "Material World: Global Family Portrait" \cite{menzel2021global} portrays families from 30 countries surrounded by their possessions, highlighting economic and cultural contexts. In "Hungry Planet: What the World Eats" \cite{menzel2021daily}, Menzel and Faith D’Aluisio showcase families with a week’s worth of food, offering insights into dietary habits and cultural preferences. The "What I Eat" project further explores this concept at the individual level, presenting daily food consumption across different cultures and professions \cite{whatieat2021}.

These projects, along with the ``Dollar Street'' initiative \cite{dollarstreet2021}, which uses photography to depict families of varying economic levels worldwide, collectively promote a more nuanced perspective on cultural diversity. ``Dollar Street'' particularly emphasizes commonalities and differences in everyday life, providing a powerful visual narrative with the potential to foster deeper understanding and empathy among viewers. \\[1em]
\textbf{Empathy and Immersive Technology:} Technological advancements in immersive technologies like VR have shown potential in fostering empathy among users from diverse backgrounds. Often described as an ``empathy machine,'' VR enhances social and emotional learning by immersing users in others' perspectives, promoting empathy through personal experiences \cite{walker2019}. Video games such as ``Detroit: Become Human'' and ``The Walking Dead'' have been studied for their ability to cultivate empathy in players, providing insights into leveraging media for empathetic development \cite{pallavicini2020}. Additionally, educational technology effectively develops empathy, especially in teaching social skills to children with Autism Spectrum Disorders \cite{daily2008}.

\subsection{XR Applications in Education and Design}

\textbf{XR in Cultural Education:}
XR technologies have moved from niche markets to broader public domains due to rapid hardware and software advancements, enhancing accessibility and effectiveness for cultural and artistic practices \cite{kolotvina2021}. These technologies have shown potential in improving educational outcomes, particularly in cultural education, by facilitating immersive learning experiences that enhance engagement and understanding \cite{al2023exploration}. Papanastasiou et al. \cite{papanastasiou2018} emphasize that XR enhances learning outcomes through interactive engagement with global cultures. Applications such as virtual tours of historical sites and interactive experiences with cultural artifacts deliver visually engaging cultural education, as noted by Huang et al. \cite{huang2019}. \\[1em]
\textbf{Design Applications:} 
In the field of interior and architectural design, VR and AR are reforming how designers conceptualize and execute their ideas by providing the experience of space before it is built \cite{racz2018vr} or interactively modifying furniture placement, style, and look \cite{moparthi2020usage}.
Moreover, this immersive interaction assists designers in better integrating diverse cultural aesthetics into their projects, as noted by Phan \& Choo \cite{phan2010}. 
Marker-less AR systems 
enable designers to experiment with different cultural themes and decorations in real-time without the need for physical markers or setups, enhancing the design process considerably \cite{johri2021}.
These applications not only enhance the design process 
but also 
help in having the final outputs closely aligned with the desired cultural aesthetics.

\subsection{Challenges and Opportunities}

\textbf{Technological Limitations:}
While offering significant potential, XR technologies also face challenges. One major hurdle is accessibility and usability; XR often requires sophisticated hardware and there is an ongoing need for research and development to make these technologies more accessible and inclusive \cite{elor2021accessibility}.
Additionally, issues with user-friendliness and a steep learning curve can further limit their widespread adoption \cite{mills2022}. Another critical challenge is achieving a high level of realism in XR environments.
Limitations in current display technology, such as resolution and field of view, can break the sense of presence and diminish the immersive experience \cite{zhan2020}. \\[1em]
\textbf{Cultural Sensitivity:}
Ensuring cultural accuracy and sensitivity in XR applications poses unique challenges. The accuracy of representation is paramount, as
misrepresentations can lead to cultural misunderstandings \cite{häkkilä2020}. Additionally, designers must navigate complex cultural contexts to avoid reinforcing existing cultural inequalities,
which requires a deep understanding of the cultural dimensions involved in design processes. As a result, cultural sensitivity becomes a crucial aspect of XR development \cite{rice2002}.

\begin{figure*}[t]
    \centering
    \includegraphics[width=\linewidth]{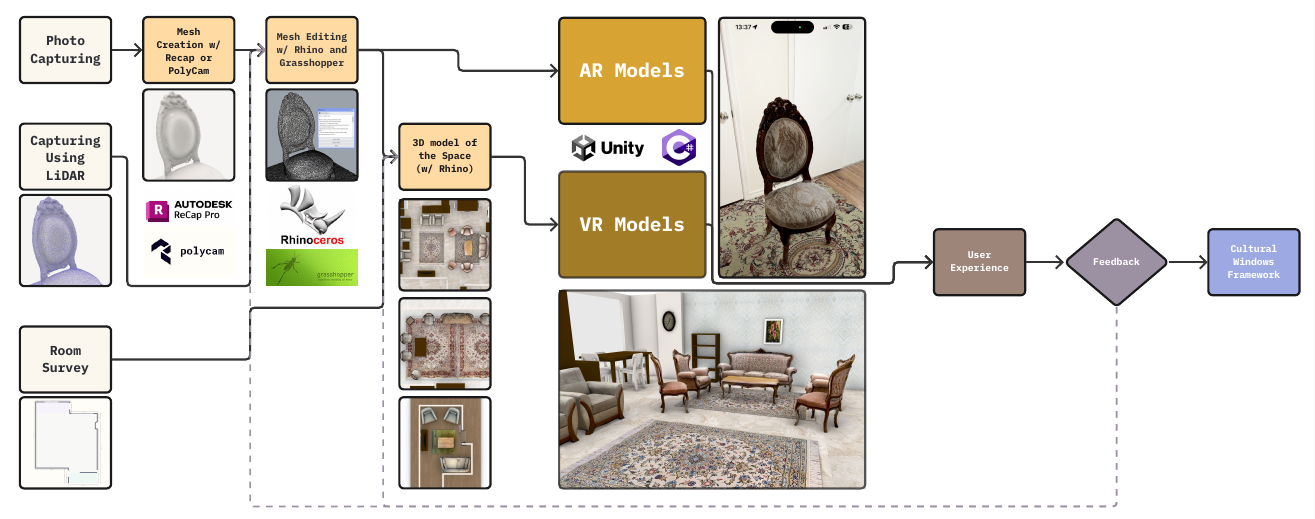}
    \caption{Methodology diagram of ``Cultural Windows,'' detailing the steps taken from data collection and mesh generation to the final user interactions with AR and VR visualizations of living room spaces.}

    \label{fig:diagram}
\end{figure*}

\subsection{Gaps in Current Research}

\textbf{Need for Integrated Approaches:}
Despite the advancements in XR technologies and their proven efficacy in various domains, there is a need for more integrated approaches that combine XR with cultural education to enhance understanding and reduce biases. Current research often treats technology and cultural content as separate entities, rather than interlinked components of a holistic educational experience.  
\\[1em]
\textbf{Potential of XR in Cross-Cultural Studies:}
Despite having examples of XR applications in cultural education, the potential of XR to provide engaging cross-cultural experiences remains underexplored. 
By providing immersive experiences of cultural simulations, XR could play a pivotal role in facilitating meaningful connections between diverse cultural groups, thus addressing one of the most critical challenges in today’s globalized society.

These gaps underscore the need for further research in integrating XR technologies with cultural education efforts, aiming to harness their potential to bridge cultural divides and enhance global understanding.


\section{Methodology}

This section details each step of our workflow, from the initial 3D scanning of real-world environments to the development and implementation of immersive VR and AR experiences (see Figure \ref{fig:diagram}). 
By outlining the methods used to collect, process, and present data, we provide a framework that can be replicated or adapted for similar studies aiming to explore the impact of immersive technologies on cultural perception.

\subsection{Site Selection and Privacy Concerns}
\label{subsec:privacy}

A core principle of ``Cultural Windows'' is to collect data such that homeowners remain in control of the data shared and that their privacy is respected.\\[1em]
\textbf{Selection of Spaces:} In the preliminary phase of this project, the selection of living spaces was intentionally limited to a few specific options to facilitate a focused and manageable setup, allowing for the evaluation and refinement of the framework. 
For this initial testing, three living room spaces were chosen from middle-class families in Tehran and Tabriz, two culturally rich cities in Iran. These spaces offered a range of design features that could demonstrate the framework's capability to capture and convey cultural nuances in an immersive virtual environment. Moreover, we considered these spaces particularly interesting as they represent locations not easily accessible to many people, enhancing their value for the ``Cultural Windows'' project.

The selected living rooms collectively exhibit a variety of design elements reflective of their cultural settings, showcasing color schemes that create unique atmospheres with traditional and contemporary aesthetic preferences, a mix of classical and modern furniture styles, and decorative elements including traditional and contemporary art, textiles, and other decor items that add depth and context to each living environment.

\begin{figure*}[t]
  \centering
   \includegraphics[width=1\linewidth]{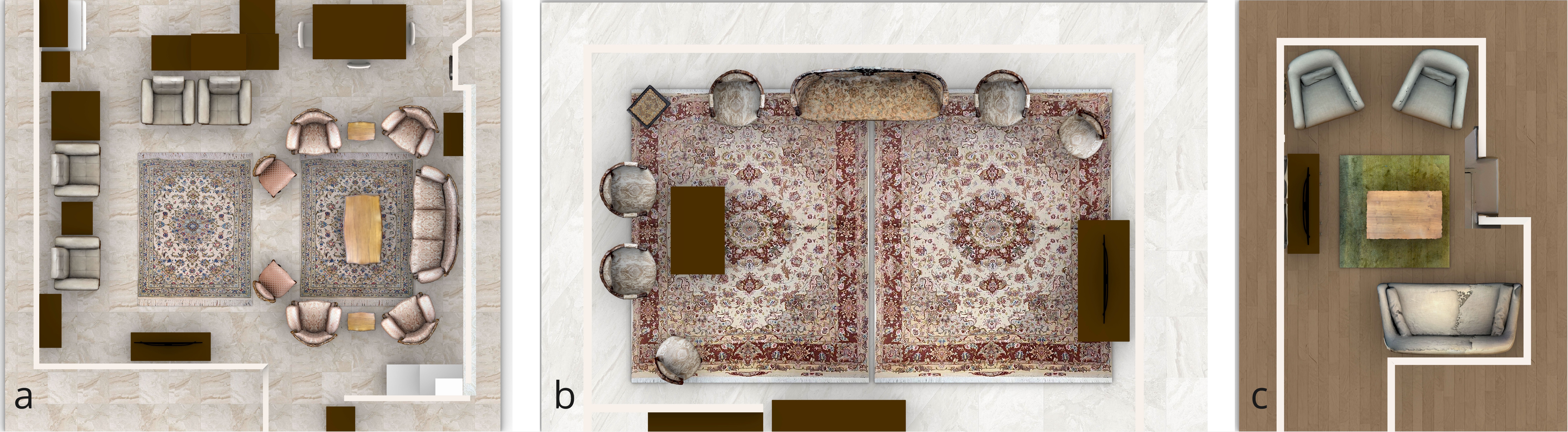}
   \caption{Plan of the three selected living rooms: (a) A living room in Tehran with a mix of modern and classical elements, (b) A living room in Tabriz with classical furniture and rugs, and (c) A cozy, modern living room in Tehran.}
   \label{fig:three}
\end{figure*}

The diversity in color, furniture, and decoration across the selected living rooms enables the project to analyze how well the ``Cultural Windows'' framework translates these design details into virtual environments that are engaging, educational, and culturally informative (See Figure \ref{fig:three}). Another goal for this deliberate selection is understanding the necessary enhancements and expansions for future iterations of the work, where a wider variety of living spaces will be integrated to increase the framework’s inclusivity and applicability across different cultural contexts.\\[1em]

\noindent\textbf{Addressing Privacy Concerns:} As previously discussed, a primary objective of ``Cultural Windows'' is to protect the privacy of homeowners willing to share their living spaces, while preserving the cultural authenticity and design of the spaces for immersive experiences. To this end, we implemented several strategies for data collection, 3D scanning, and modeling. Homeowners were invited to selectively share detailed scans or photographs of textures or artworks that they were comfortable disclosing and in their opinion, had a cultural or personal significance for them. Through this approach, we aimed to address both privacy concerns and technical challenges associated with scanning entire rooms. In the next section of the methodology, we will discuss possible 3D scanning and modeling approaches, and our proposed method to protect homeowners' privacy (see \ref{subsec:scanning}).

\subsection{3D Scanning Approaches}
\label{subsec:scanning}

\begin{figure}[H]
  \centering
   \includegraphics[width=1\columnwidth]{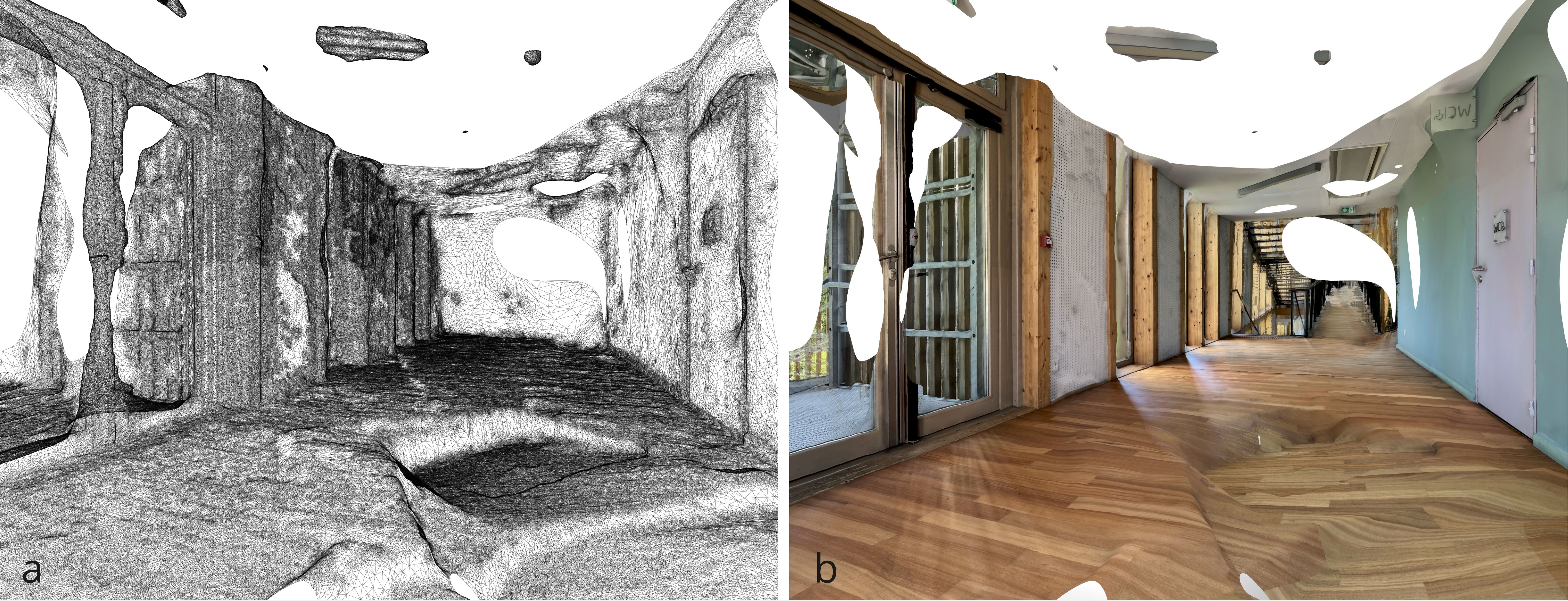}
   \caption{Reconstruction of entire interior spaces often reveals photogrammetry limitations: (a) A mesh generated from ~150 photos using Autodesk ReCap Photo, (b) A rendered view of the same space.}
   
   \label{fig:recap}
\end{figure}

A wide variety of different 3D scanning and reconstruction approaches exist~\cite{ma2018review}, many of which rely on specialized equipment, such as commercial 3D scanners or depth-cameras using RGB-D or time-of-flight technologies. For the methodology presented here, we aimed for a widely accessible approach, in particular, we focused on the use of cameras included in mobile phones and 3D scanning apps making use of these cameras to create 3D models. Consequently, our workflow only considers photogrammetry, which can be used with almost any phone that includes a camera, and LiDAR, which arguably is only available in high-end models.
In this section, we first describe the 3D scanning technologies and approaches used in this project, discussing their advantages and limitations. 
We then present our proposed hybrid scanning and modeling method to create 3D models of living spaces, addressing both the technological challenges and privacy goals of ``Cultural Windows.'' \\[1em] 
\textbf{Photogrammetry} was employed using a smartphone camera, in our case, an iPhone, to capture detailed textures and geometries of the selected living spaces. The number of required photos was different based on the size and details of objects or spaces and ranged from 50 to 100. Photos were processed using Autodesk Recap Photo to construct 3D meshes \cite{autodesk2023}. While photogrammetry is generally effective, it can sometimes miss fine details and produce meshes with washed-out textures or incomplete data, especially when scanning a large interior space (See Figure \ref{fig:recap}).

\begin{figure}[H]
  \centering
   \includegraphics[width=1\columnwidth]{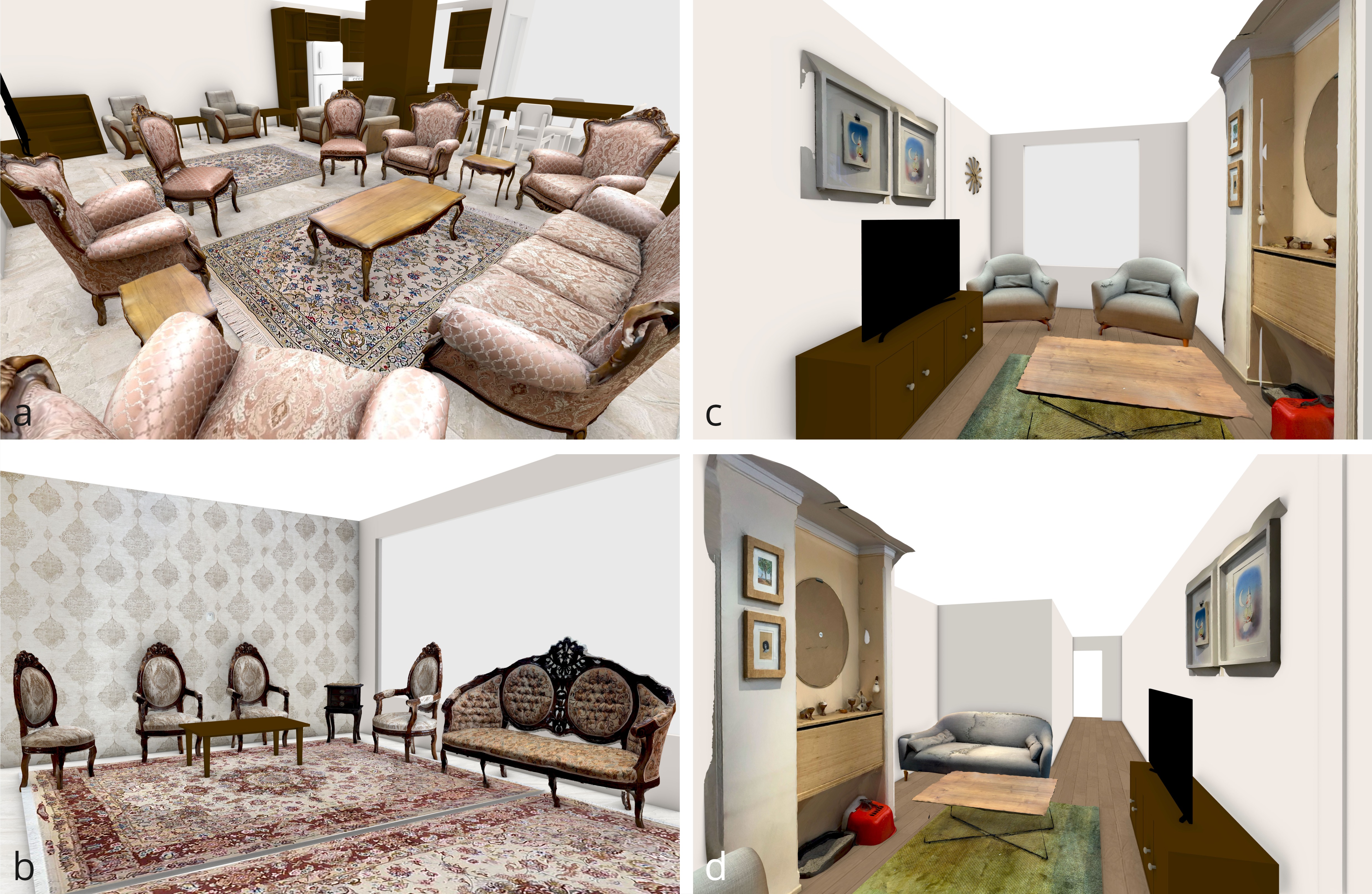}
   \caption{Final reconstructed 3D models of the living spaces, ready for VR integration: (a) Living room No. 1 in Tehran, (b) Living room No. 2 in Tabriz, (c) and (d) Living room No. 3 in Tehran.}
   
   \label{fig:all}
\end{figure}


\begin{figure*}[t]
    \centering
    \includegraphics[width=\linewidth]{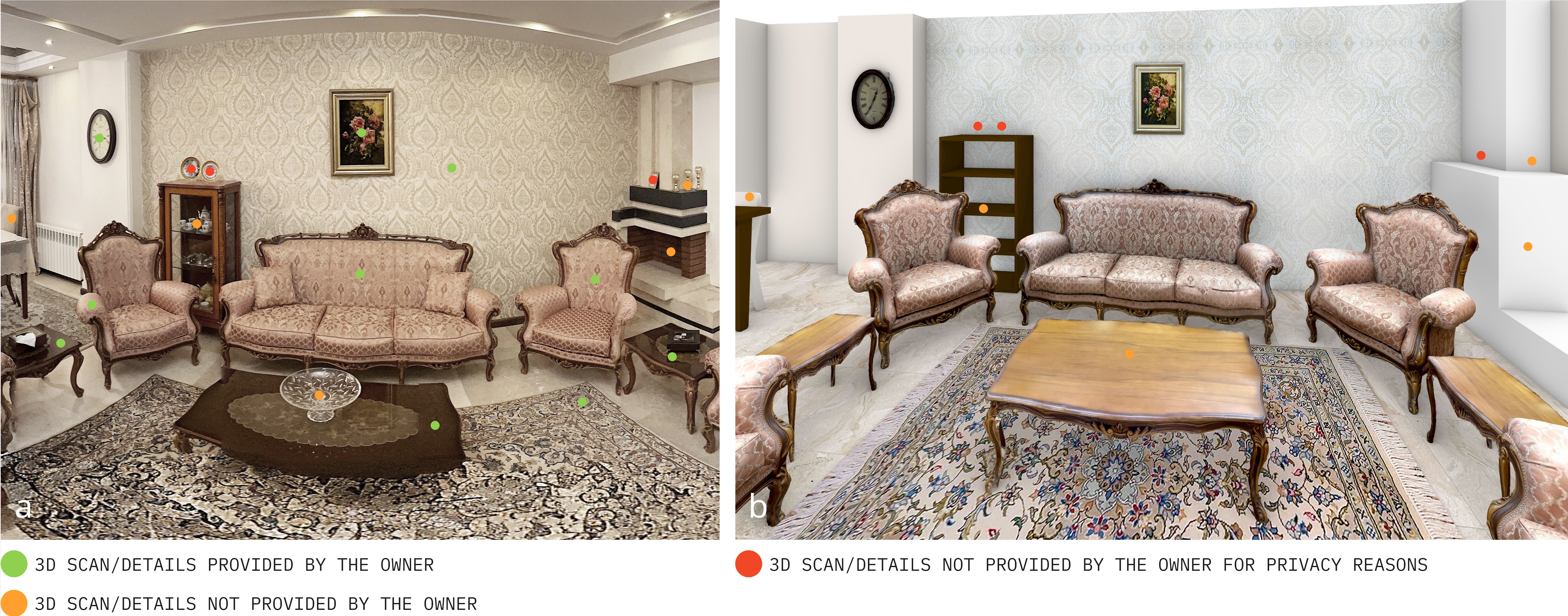}
    \caption{A side-by-side comparison of (a) the actual living space, and (b) the reconstructed 3D model using our framework. Green, orange, and red dots indicate design elements that were shared by the owner and those that were not, with unshared elements replaced by generic objects and models in our reconstructed version.}

    \label{fig:compare}
\end{figure*}

\noindent\textbf{LiDAR scanning}, performed with the LiDAR sensor available on newer iPhones, was conducted using the PolyCam app to capture precise spatial layouts and larger structural elements of the rooms \cite{polycam2023}. LiDAR scanning can enhance accuracy in spatial data acquisition, thus enhancing the overall model quality. However, in cases where a LiDAR sensor is not available, photogrammetry alone has proven sufficient for capturing necessary details. \\[1em]
\textbf{Hybrid Scanning and Modeling Approach:} To address the technical and privacy challenges associated with 3D scanning entire spaces, we propose a hybrid scanning and modeling approach. In this method, homeowners provided general dimensions, hand-sketched plans, or any other details they were comfortable sharing. These details were used to create a non-photorealistic version of the space with generic objects (see Figure \ref{fig:teaser}b). When technically feasible and when owners are comfortable with the privacy implications, the ``Room'' feature of PolyCam was used to reconstruct 3D models of the spaces without compromising personal details. The following section will elaborate on the subsequent steps to develop 3D models and immersive experiences for ``Cultural Windows.''

\subsection{Modeling and XR Development}
\label{subsec:models}

In this section, we provide a detailed walkthrough of the steps taken after scanning individual objects and obtaining the necessary details from owners to create 3D models of living spaces. This preparation is crucial for developing detailed 3D models that can later be used to create immersive experiences.\\[1em]
\textbf{Data Processing and Mesh Reconstruction:} After collecting the raw data for individual furniture and cultural elements using photogrammetry and LiDAR and generating initial meshes with Autodesk Recap Photo and PolyCam, the next step in our workflow involves refining these into detailed 3D models. We use mesh editing tools in Rhino 3D \cite{rhino2023} to repair meshes (e.g., filling holes) and ensure all objects are constructed from error-free, usable meshes, as illustrated in Figure \ref{fig:diagram}. Additionally, we ensure texture mapping and material assignment is done properly to create realistic 3D scenes of each living space. Once all objects are processed and ready, we proceed to prepare the detailed 3D models of each living space.

\begin{figure}[H]
  \centering
   \includegraphics[width=1\columnwidth]{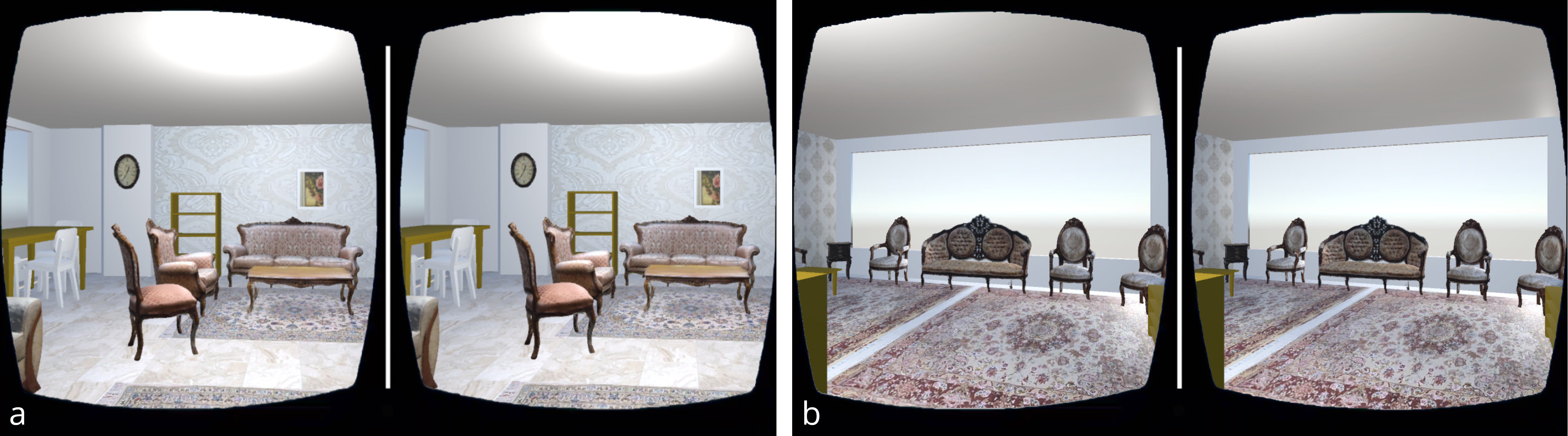}
   \caption{Stereoscopic screenshots showing (a) Living room No. 1 in Tehran, and (b) Living room No. 2 in Tabriz.}
   \label{fig:VR}
\end{figure}

\noindent\textbf{Reconstruction of the Spaces:} Based on the details provided by users for the overall model of their living space, the non-photorealistic 3D model for the basic structure or furniture/design elements not shared by homeowners can be created. If users only provide general dimensions, hand sketches, etc., we manually model the space in modeling platforms. When using the ``Room'' feature of PolyCam, the 3D model generated is imported into a 3D modeling environment. The next step involves importing the detailed 3D models of culturally significant objects, rendered with original materials and textures in a photorealistic style, into these generic models. To accurately place each object, we refer to the notes and explanations provided by users. Figure \ref{fig:all} illustrates the final reconstructed 3D models of the living spaces. Once the models are created, 3D renders of the main views of the interior spaces are sent to owners for review and approval. This ensures that we meet our privacy goals and only share what owners agree to share. Figure \ref{fig:compare} demonstrates this approach, contrasting it with an actual photograph of one of the spaces, which the owner agreed to share. It highlights objects omitted by the owner due to privacy concerns, perceived cultural insignificance, or technical scanning difficulties. In this project, all modeling and integration tasks were performed using Rhino 3D. \\[1em]
\textbf{VR and AR Implementation:} The Next step in the workflow is the development of immersive VR and AR experiences. Using both the 3D models for individual objects and the overall scenes representing interior spaces, VR and AR experiences are developed to allow users to explore the cultural design elements and the living spaces in an interactive manner. The XR development was carried out using Unity, which provides robust support for both VR and AR applications \cite{unity2023}. For the VR implementation, we used the Oculus Meta Quest 2 \cite{oculusmetaquest2023} headset to provide the immersive experience of exploring the replica of the living spaces (See Figure \ref{fig:VR}).

\begin{figure}[H]
  \centering
   \includegraphics[width=1\columnwidth]{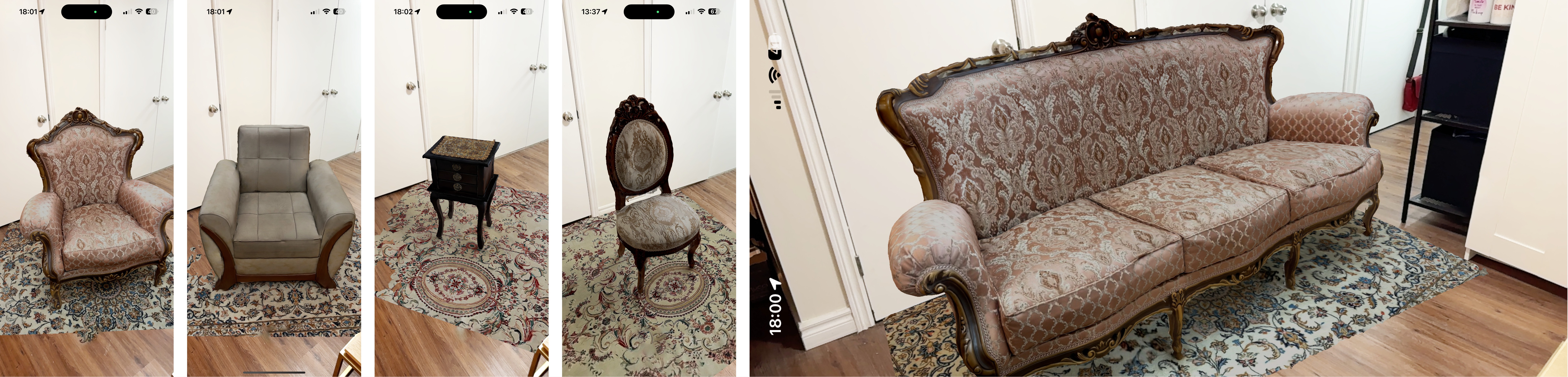}
   \caption{Users can use a cellphone or tablet to superimpose 3D scanned cultural elements, such as traditional furniture, into their own living room space in AR mode. This allows for a comparison of cultural elements with their existing domestic designs.}
   \label{fig:AR}
\end{figure}

In AR, users see digital overlays of cultural objects and design elements, such as traditional furniture and rugs, superimposed onto their real-world environment (See Figure \ref{fig:AR}). 

This method enabled users to see and interact with virtual representations of culturally significant objects within spaces familiar to them using their mobile devices, allowing for a direct comparison between local and foreign cultural elements. This is a critical aspect of the work as it enhances the educational aspect of the project by embedding global culture into a personal context. \\[1em]
\textbf{User Interaction Design:}
The design of user interactions in both VR and AR aims to enhance educational outcomes and cultural understanding. In VR, basic navigation controls enable users to move freely around the virtual spaces, facilitating the exploration of different cultural environments.

In AR, users select from a curated list of furniture and design items—derived from detailed 3D scans and mesh editing processes as described in \ref{subsec:scanning}—and integrate them into spaces familiar to them. This functionality allows users to position, scale, and orient cultural objects alongside their existing furnishings, enabling them to experiment with and visualize how diverse cultural styles can harmonize with their personal environments. This hands-on approach has the potential to inform users about global design aesthetics and inspire them to incorporate these elements into their own spaces, providing a practical demonstration of cultural adaptation.

\begin{table*}[t]
\centering
\caption{Participant Demographics and Professions}
\label{tab:participant_info}
\begin{tabularx}{\textwidth}{@{}XXXl@{}}
\specialrule{1.5pt}{0pt}{0pt} 
\textbf{Participant} & \textbf{Referred to As} & \textbf{Background} & \textbf{Profession} \\ 
\midrule
Participant Male 1   & PM1                  & Italian             & Senior Architect    \\ 
Participant Male 2   & PM2                  & Chinese             & Senior Interior Designer    \\ 
Participant Female 1 & PF1                 & Moroccan            & Intermediate Designer and Educator \\ 
Participant Female 2 & PF2                 & American            & Junior Interior Designer \\ 
\specialrule{1.5pt}{0pt}{0pt} 
\end{tabularx}
\end{table*}

\subsection{Preliminary Informal Study}

The preliminary informal study conducted as part of the ``Cultural Windows'' project was designed to capture initial perceptions and explore the potential of immersive technologies in enhancing cultural understanding. This study focused on interior design and architecture professionals, using their expertise to gain insights into how such individuals perceive and interpret cultural spaces. This study aimed to iteratively refine the methodology by understanding what stands out to these professionals and how the methodology could be improved while adhering to principles of privacy and selective inclusion of objects and furniture. \\[1em]
\textbf{Participant Selection:}
For this preliminary study, we asked 4 participants to participate in this project. Participants were selected based on their professional background in interior design and architecture, ensuring they had the necessary expertise to critically describe living spaces and analyze the outcome of immersive experiences on their thoughts and design approach. This professional focus was chosen because these individuals are trained to notice and consider aspects of space and design that might be overlooked by the general public, providing depth to the analysis of cultural representation in living spaces. The participant group consisted of two men and two women, each from different cultural environments to facilitate a broader exploration of how cultural influences shape perceptions of unfamiliar living spaces (See Table \ref{tab:participant_info}). 

While the study was led by an individual familiar with Iranian culture, all participants were unfamiliar with the culture under study--Iranian middle-class living rooms; none had previously studied or visited the country. \\[1em]
\textbf{Study Design:}
Participants were initially asked to describe, purely from their imagination and without any visual aids, what they envisioned as a typical middle-class family living room in Iran. This exercise aimed to capture the common elements and themes that emerged from their descriptions and to examine how these elements align with actual Iranian living spaces. These descriptions were recorded through detailed note-taking.

After providing their descriptions, participants were asked to enter the study space individually to work with AR and VR applications. The study was conducted in an empty conference room, creating a neutral environment to focus on the AR and VR experiences without external influences. For the AR experience, participants used a mobile phone with an AR application to select and place furniture items from the studied living rooms into the study space. This step assessed how they envisioned integrating various design elements into a real-world context.

Subsequently, participants were equipped with a VR headset to explore the three different living rooms, as discussed in \ref{subsec:models}. This immersive experience aimed to provide a deeper understanding of how cultural design elements were perceived. \\[1em]
\textbf{Feedback Collection:}
Feedback was collected from participants after their AR and VR experiences to understand how their perceptions changed through these immersive interactions. Participants were asked to reflect on the accuracy of their initial descriptions compared to the VR environments and to comment on the effectiveness of this immersive approach for different applications, such as education, professional training, and cultural appreciation. The feedback was gathered through note-taking to ensure that the study remained informal and iterative, focusing on methodological improvements rather than formal research outcomes.


\section{Observations} 
\label{sec:analysis}

In this section, we present a qualitative analysis of feedback from participants who engaged with AR and VR technologies to explore cultural differences in living space design. We examined their responses to understand how these immersive tools influenced their perceptions of cultural elements in interior design. The analysis synthesizes insights from participants' initial descriptions, their interactions with AR and VR environments, and their reflections on the educational value and cultural accuracy of these experiences. By integrating direct quotations and thematic content analysis, we provide a nuanced understanding of the impact of these technologies on cultural education. This approach helps identify key themes and draw conclusions about the effectiveness of AR and VR in enhancing cross-cultural understanding and design acumen.

\subsection{Cultural Perception and Initial Descriptions}

Participants began the study with distinct perspectives shaped by their individual cultural backgrounds and understanding of Iranian culture. Their descriptions incorporated a variety of design elements they associated with Iranian living spaces, such as ``colorful textiles'' and ``communal seating arrangements'' (\textit{PM1}). As \textit{PM2} mentioned, his inclusion of ``a mix of seating options... reflects a typical family-oriented space,'' illustrating a blend of functionality with cultural aesthetics. These initial descriptions, while rich and imaginative, varied in accuracy, reflecting participants' diverse levels of exposure to and understanding of the actual cultural context.

\subsection{Challenges and Risks}

After discussing the initial cultural perceptions, it is important to address the inherent challenges and risks early on. This provides a clear understanding of the limitations and ethical considerations that frame the subsequent exploration of AR and VR technologies.

In assessing the challenges and risks of experiencing cultural contexts through AR or VR, participants reflected on the potential of technologies to provide insights into an unknown culture. While the experiences were generally seen as detailed and culturally engaging, concerns were raised about the depth and authenticity of these representations.

\textit{PM1} highlighted that while the traditional elements felt authentic, there's a continuous need for collaboration with cultural experts to ensure that modern influences and subtler cultural nuances are accurately depicted. This is crucial to avoid oversimplifications that could lead to misrepresentations. 
He pointed out, ``While the traditional elements felt authentic, it's crucial to continue collaborating with Iranian cultural experts to ensure all aspects, including the modern influences, are accurately depicted.''

\textit{PM2} appreciated the overall accuracy but worried about a potential for interpretation biases that technology might introduce. He suggested that for the development of this framework integrating validations from native cultural insiders could enhance credibility, stating, ``It will be beneficial to have some sort of validation from natives within the experience, perhaps through testimonials or cultural notes, to affirm the authenticity'' of what being shown to participants. 

\textit{PF1} and \textit{PF2} both acknowledged their limited ability to judge the full accuracy of the cultural representation due to their non-Iranian backgrounds. Similar to \textit{PM2}, they also emphasized the importance of involving people who are deeply familiar with the culture in the creation and review processes. \textit{PF1} suggested, ``I would appreciate more direct input from Iranian cultural historians or artists as part of the experience to ensure its accuracy,'' emphasizing the need for expert involvement to enhance the educational quality and authenticity of the representations.

\textit{PF2} expressed a need for a more structured feedback mechanism within the VR and AR platforms to ensure that any cultural content is vetted thoroughly. She proposed, ``Incorporating feedback mechanisms where users can learn about the development process and the sources used for the cultural elements might help ensure greater authenticity.''

\subsection{Impact of AR on Cultural Integration}

The AR experience helped in merging \textit{PF2}'s abstract concepts with tangible cultural elements. She noted, ``I was surprised by how the traditional furniture felt more in place with the colorful and welcoming atmosphere I imagined.'' This technology allowed participants to experiment with placing cultural items within a familiar environment to help them address some misconceptions and possibly deepen their appreciation for Iranian contemporary design principles. \textit{PM1}'s experience where ``some traditional pieces... integrated seamlessly'' with his descriptions highlighted the AR's capability to enhance cultural authenticity in a familiar setting.

\subsection{VR Experience and Cultural Immersion}

VR experiences expanded cultural insights for most of the participants by immersing them in recreated Iranian living rooms. \textit{PM1} found the VR environment ``revelatory,'' as it combined traditional elements he anticipated with modern designs he had not considered. The VR setup provided a ``more dynamic and detailed'' view (\textit{PM1}), assisting in a better understanding of how traditional and contemporary styles can coexist. \textit{PF2} expressed how the VR environments were ``eye-opening and beautifully crafted,'' and \textit{PM2} mentioned that by showing ``cultural aesthetics, from intricate carpet designs to the layout optimized for family gatherings,'' ``the VR settings... provided a deeper understanding of the actual cultural context.''

\subsection{Educational Values}

Three of the participants agreed on the potentially high educational value of using AR and VR technologies. \textit{PM1} highlighted that ``VR technology can be a powerful tool in my work'' and ``incredibly beneficial in educational settings,'' emphasizing its potential in professional architectural practices and education. The immersive nature of these technologies was particularly praised for their ability to convey complex cultural and design concepts interactively. \textit{PF1} saw VR as ``an exceptional educational tool, especially in cultural studies or art history'', where the immersive experience could transform abstract cultural concepts into tangible, relatable experiences.

\subsection{Suggestions for Improvement}

Participants suggested specific improvements to increase the educational and professional use of our AR and VR representations. \textit{PM2} recommended ``increasing the number of detailed models and cultural elements in the scenes,'' emphasizing the need for capturing more design and decoration elements and incorporating them into our models. \textit{PF1} called for ``a wider range of cultural scenarios,'' suggesting that inclusivity could enhance the representativeness and educational quality of these experiences. One other important feedback for improvement was focused on providing specific information about the cultures and design elements through interactive methods in VR and AR. Examples are: ``Integrating more contextual information, like historical and cultural backgrounds of the designs'' (\textit{PM2}) or ``Adding guided tours or expert commentary within the VR experiences'' to enhance understanding (\textit{PF1}).


\section{Discussion}
\label{discussion}

This informal preliminary study gathered insights from design professionals through AR and VR experiences to evaluate the potential for enhancing the understanding and appreciation of cultural differences in living space design via a newly proposed framework for creating immersive cultural experiences. The goal was to collect their feedback and insights to iterate on our methodology, guiding further refinement to better align with the project's objectives of privacy, cultural sensitivity, and technological feasibility.

Consistent with prior research, our findings highlight that immersive technologies can deepen users' engagement and enhance their understanding of complex concepts such as cultural heritage and spatial aesthetics \cite{bekele2019, komianos2022, carrozzino2010}. Similar studies have demonstrated that VR, in particular, offers opportunities for users to experience and interact with cultural elements in ways that are not possible through traditional educational methods \cite{young2020}. This preliminary study extends the existing literature by showing how AR complements VR, enabling users to visualize and personalize the integration of foreign cultural elements into familiar environments within the proposed framework. 

Our preliminary study supports the theoretical framework that suggests immersive technologies can bridge the gap between abstract knowledge and experiential understanding \cite{oluseye2020}. By immersing participants in VR environments that are contextually accurate, and by overlaying AR elements onto participants' real-world settings, the technologies have the potential to promote a more nuanced understanding of cultural differences. This aligns with the constructivist learning theory, which posits that learners construct knowledge best through experiential, hands-on activities \cite{piaget1954}.

Practically, this study underscores the potential of AR and VR in professional fields related to design and architecture, especially within the proposed framework to create diverse cultural experiences. Participants noted that these technologies could positively impact how cultural sensitivities are integrated into design processes. This is particularly relevant in today's globalized market, where understanding and incorporating diverse cultural aesthetics are crucial for the success and relevance of design projects.

A key finding from this study is the critical role of cultural authenticity and sensitivity in the development of AR and VR content within the proposed framework. Participants valued experiences that were not only immersive but also culturally accurate and sensitive. This emphasizes the need for developers of educational content in AR and VR to work closely with cultural experts to ensure that representations do not perpetuate stereotypes or misrepresentations. Additionally, participants highlighted the importance of trust in the cultural accuracy of the immersive experiences. Knowing that the content is developed or validated by cultural experts can enhance users' trust and confidence in the authenticity of the representations. This underscores the need for a validation process involving native cultural insiders to ensure the credibility of the immersive experiences, thereby increasing their educational and empathetic impact.


\section{Limitations}

This study, while providing valuable insights into the use of AR and VR technologies in cultural understanding and education, is subject to several limitations that must be acknowledged. First, the sample size of the study was very small and only consisted of professionals within specific fields related to architecture and design. While this provided targeted insights, it limits the generalizability of the findings to broader populations or to individuals with different professional backgrounds or cultural expertise. Additionally, all participants had a similar, limited familiarity with the culture under study. Future research could explore the impacts of varying levels of cultural familiarity and pre-existing biases on the perceived authenticity and effectiveness of the experiences, further connecting these factors with the authenticity findings from this study.

Furthermore, the AR and VR experiences used in this study were limited to specific cultural elements and a limited number of living space designs from Iranian culture. While this focus provided some insights into one cultural context, it may not encapsulate the diversity and complexity of the designs within the studied context and, more importantly, the global cultural designs. 

Another limitation concerns the technological aspects of the AR and VR systems used, as well as the challenges associated with data collection and the gathering of 3D scans from different cultures and locations. These systems, while state-of-the-art, still have inherent limitations in terms of resolution, user interface, and interactivity that might affect the immersion and realism of the experiences. Specifically, the process of collecting 3D scans, as evidenced by the third living room space in Tehran, illustrates additional technological hurdles. Despite the accessibility of smartphones with reliable cameras, factors such as the expertise of the individual capturing the images, varying light conditions, and camera resolution can significantly impact the quality of the scans. In this instance, the quality of the scans was noticeably lower compared to other examples, which may be due to a lack of expertise or other environmental variables (See Figure \ref{fig:all} (c) and (d)). Additionally, issues such as the comfort and accessibility of VR headsets can influence user engagement and the overall effectiveness of the experiences, further complicating the deployment of these technologies in diverse settings.


\section{Future Research Directions}

To address the limitations of the proposed framework and build on its current state future research should aim to expand the study to include various examples from different socio-economic classes of living within each culture and multiple cultural contexts, and employ a mix of quantitative and qualitative research methods to provide a deeper understanding of the use of ``Cultural Windows'' together with immersive technology.

Enhanced technological aspects of AR and VR systems, such as improved user interfaces with the possibility of providing detailed information about specific design elements in each culture, interactivity, and realism, will further the immersion and educational impact of these experiences.

The incorporation of rigorous validation processes, involving cultural experts and native individuals in the content creation and review phases, will be crucial for any ``Cultural Window'' made publicly available as a reliable window into a different culture. This will ensure that the cultural representations in VR and AR are accurate and sensitive, avoiding potential misrepresentations and fostering a more inclusive and respectful global understanding.

\subsection{Collaborative and Automated Future}

What we envision for the future of ``Cultural Windows'' is a network of systems that facilitates data collection from various cultures and locations. Such a system would enable users globally to enhance ``Cultural Windows'' by uploading 3D scans or photos of culturally valuable artifacts from their living spaces to a shared platform. Subsequent automated processes could then generate and refine detailed 3D models of these significant items and the living spaces. This approach would not only democratize the contribution and sharing of cultural knowledge but also enhance the richness and authenticity of the educational content available in AR and VR platforms.

The envisioned network of systems represents a significant advancement in cultural education technology. By automating the modeling process and facilitating widespread contributions, this system promises to revolutionize how cultural heritage is preserved, studied, and appreciated. Designers, educators, and curious individuals will have unprecedented access to a repository of immersive experiences that are not only educationally valuable but also culturally enriching.


\section{Conclusion}
\label{sec:conclusion}

By introducing a framework to create immersive cultural experiences, we contribute a mechanism to use immersive technologies to foster cross-cultural understanding. In our preliminary study, most participants noted the practical applications of projects like ``Cultural Windows'' in their professional fields and the educational value of such immersive experiences.

By further embracing collaborative contributions and advancing technological solutions, future work can transform how we experience, understand, and appreciate global cultures through immersive technologies.


\section*{Supplemental Materials}
\label{sec:supplemental_materials}

All supplemental materials are available on \href{ https://osf.io/rvqxa/?view_only=4f63213cb3314472a570a352a5b022f8}{OSF} (\url{https://t.ly/D2K3A}). The repository includes the Creative Commons version of all figures in this article, the descriptions provided by participants in the first phase, and the questionnaire used during the debriefing phase after the immersive experience.


\section*{Acknowledgments}

This research is partially funded by the MOPGA program, implemented by Campus France, and funded by the French Ministry for Europe and Foreign Affairs in collaboration with the French Ministry for Higher Education and Research. It was also supported by the Agence Nationale de la Recherche (ANR), grant number ANR-19-CE33-0012.


\bibliographystyle{plain}
\bibliography{references}

\end{multicols}

\end{document}